# Realization of Gain with Electromagnetically Induced Transparency System with Non-degenerate Zeeman Sublevels in $^{87}$Rb

**Minchuan Zhou,[1] Zifan Zhou,[2] and Selim M. Shahriar[1,2,*]**

[1]*Department of Physics and Astronomy, Northwestern University, Evanston, IL 60208, USA*

[2]*Department of EECS, Northwestern University, Evanston, IL 60208, USA*

*Email: minchuanzhou2013@u.northwestern.edu

**Abstract** Previously, we had proposed an optically-pumped five-level Gain EIT (GEIT) system, which has a transparency dip superimposed on a gain profile and exhibits a negative dispersion suitable for the white-light-cavity signal-recycling (WLC-SR) scheme of the interferometeric gravitational wave detector [Phys. Rev. D. **92**, 082002 (2015)]. Using this system as the negative dispersion medium (NDM) in the WLC-SR, we get an enhancement in the quantum noise (QN) limited sensitivity-bandwidth product by a factor of ~18. Here, we show how to realize this GEIT system in a realistic platform, using non-degenerate Zeeman sublevels in alkali atoms. Specifically we choose $^{87}$Rb atoms, which produce the negative dispersion around 795nm. The current LIGO operates at 1064nm but future LIGO may operate at a wavelength that is consistent with this atomic system. We present a theoretical analysis for the susceptibilities of the system. To account for the QN from the GEIT system, it is necessary to use the master equation (ME) approach. However, due to the number of energy levels involved, applying the full ME approach to this system is very complex. We have also shown earlier, in the reference cited above, that under GEIT condition, the net enhancement in the sensitivity-bandwidth product, as predicted by the ME, is close to that predicted by applying the Caves model for a phase-insensitive linear amplifier. Therefore we here use the Caves model for the QN from the NDM and this simplified numerical model shows that the enhancement of the sensitivity-bandwidth product as high as 17 is possible.

## I. INTRODUCTION

Previously, we had presented an interferometric gravitational wave (GW) detector using a white light cavity [1,2,3,4,5,6,7] for signal recycling (the WLC-SR scheme), which can enhance the quantum noise (QN) limited sensitivity-bandwidth product [8]. The key element in the WLC is a negative dispersion medium (NDM), with vanishingly small additional noise, used to compensate the phase variation due to change in frequency, including optomechanical effects. For realizing such an NDM, we had proposed a system using five energy levels in the M-configuration that produces gain with electromagnetically induced transparency. The quantum noise from this configuration, which we call a GEIT system, was evaluated rigorously using the master equation (ME) approach [9] in our numerical simulation. The resulting sensitivity-bandwidth product is enhanced by a factor of ~18 [8] compared to the highest sensitivity result predicted by Bunanno and Chen [10]. To the best of our knowledge, such a GEIT system has not been studied to date, neither experimentally nor theoretically. As such, it may not be a-prior obvious whether such a system can be realized at all in practice. In this paper, we describe an explicit realization of the GEIT system, using non-degenerate Zeeman sublevels in alkali atoms, specifically $^{87}$Rb atoms. In Sec. II, we describe the Rb GEIT system in detail. In Sec. III, we theoretically model the GEIT system using the density-matrix approach and calculate the quantum noise limited sensitivity of the WLC-SR detector incorporating this system in the WLC. In Sec. IV, we summarize the results and present an outlook for future studies.

## II. DESCRIPTION OF THE GEIT SYSTEM USING $^{87}$Rb

We use the 17 non-degenerate Zeeman sublevels in $^{87}$Rb atoms as shown in Fig. 1 for realizing the GEIT system. The optical excitation scheme is illustrated schematically in Fig. 2(a). The five levels constituting the M-type system as illustrated in Fig. 2(b) are represented by the following Zeeman sub-levels: $|1\rangle = 5S_{1/2}, F = 2, m_F = -2$, $|2\rangle = 5S_{1/2}, F = 2, m_F = 0$, $|3\rangle = 5S_{1/2}, F = 2, m_F = 2$, $|4\rangle = 5P_{1/2}, F = 1, m_F = -1$, and $|5\rangle = 5P_{1/2}, F = 1, m_F = 1$. These states belong to only two hyperfine levels: {$5S_{1/2}$, F=2} and {$5P_{1/2}$, F=1}. However, in order to ensure that these levels produce the desired GEIT effect, it is also necessary to make use of additional hyperfine levels, namely {$5S_{1/2}$, F=1}, {$5P_{3/2}$, F=1}, and {$5P_{3/2}$, F=2}. For the parameters and conditions considered here, as explained in detail later, the effect of the remaining hyperfine levels within the $D_1$ and the $D_2$ transitions can be ignored.

In order to lift the degeneracy between the Zeeman sublevels, we assume the application of a moderate magnetic field along the quantization axis. The Lande $g_F$-factors for each of the five hyperfine levels are shown in Fig. 1. For a magnetic field strength $B$ (in Gauss), the energy shift for a Zeeman sublevel with quantum number $m_F$ is given by $1.4 g_F m_F B$ MHz. The strength of $B$ is to be kept low enough so that the Zeeman splitting between adjacent $m_F$ levels is small compared to the hyperfine splitting within the corresponding fine structure.

The transitions $|1\rangle$-$|4\rangle$, $|2\rangle$-$|4\rangle$ and $|3\rangle$-$|5\rangle$ are coupled by the pump fields $\Omega_1$ ($\sigma^+$-polarized), $\Omega_2$ ($\sigma^-$-polarized), and $\Omega_4$ ($\sigma^-$-polarized), respectively, while the transition $|2\rangle$-$|5\rangle$ is coupled by the probe field $\Omega_3$ ( $\sigma^+$-polarized). The pump fields and the probe field are all below resonance. (We assume the use of cold atoms, so that Doppler broadening is neglected.)

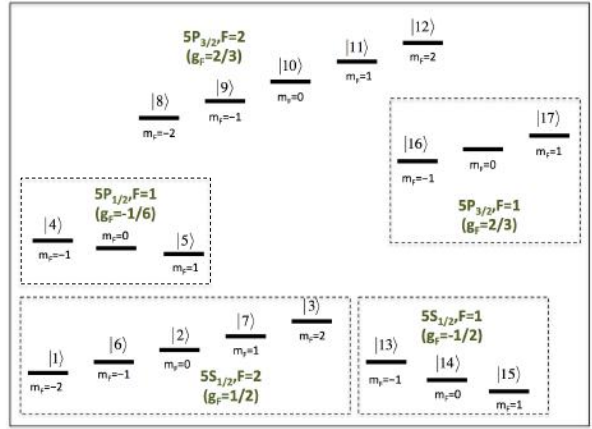

FIG. 1. 17 non-degenerate Zeeman sublevels in $^{87}$Rb atoms used for realizing the GEIT system.

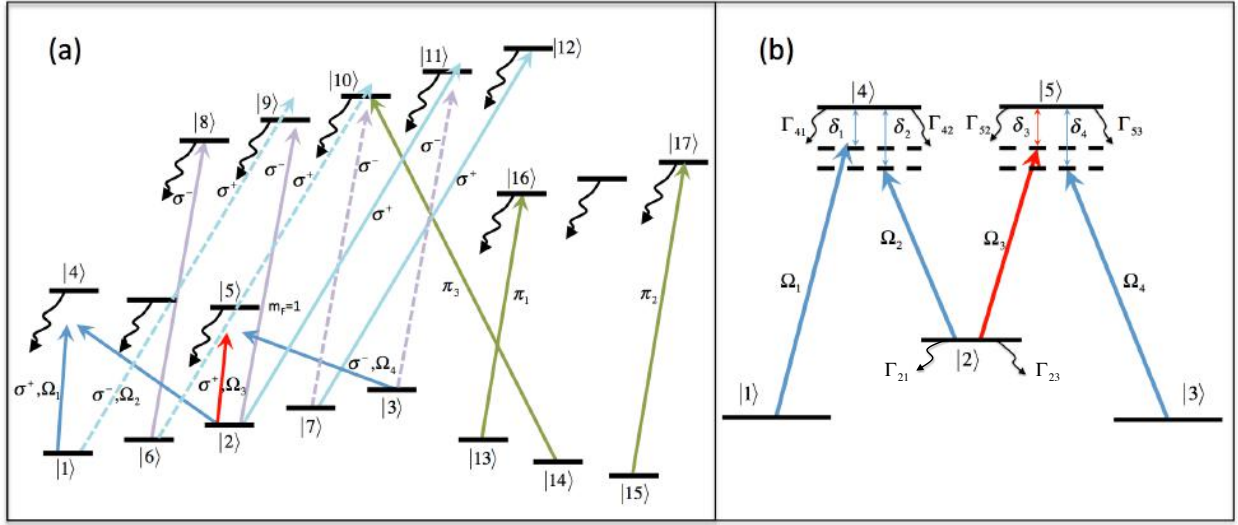

FIG. 2. (a) Illustration of the optical excitation scheme for realizing the M-type gain with electromagnetically induced transparency (GEIT) system using $^{87}$Rb atoms; (b) Schematic illustration of the effective five-level system that results from these interactions. The decay rates for state $|2\rangle$, as shown in Fig. 2(b), are due to the optical pumping via its coupling to the 5P$_{3/2}$, F=2 hyperfine state.

The optical pumping beams applied on the D2 transition are also shown in Fig. 2(a). The two $\pi$-polarized lights ($\pi_1$ and $\pi_2$) coupling the F=1,m$_F$=±1 ground states to the F=1,m$_F$=±1 states in the 5P$_{3/2}$ manifold and the $\pi$-polarized light ($\pi_3$) coupling the F=1,m$_F$=0 ground state to the F=2,m$_F$=0 state in the 5P$_{3/2}$ manifold ensure that no atoms can get trapped in the 5S$_{1/2}$, F=1 state. The $\sigma^+$-polarized and $\sigma^-$-polarized optical pumping beams ensure that atoms would not get trapped in the $5S_{1/2}$, F=2, $m_F=\pm 1$ states (levels $|6\rangle$ and $|7\rangle$). Furthermore, in the absence of the beams that excite the GEIT transitions, these optical pumping beams would send all the atoms into the ground states $|1\rangle$ and $|3\rangle$, thus producing the population imbalance necessary for Raman gain.

The detunings of the pump fields and the probe field are denoted by $\delta_j$ ($j=1,2,3,4$). The detunings of the pump fields $\Omega_1$ and $\Omega_2$ are chosen to balance the differential light shift experienced by levels $|1\rangle$ [$\Omega_1^2/(4\delta_1)$] and $|2\rangle$ [$\Omega_2^2/(4\delta_2)+\Omega_3^2/(4\delta_3)$], so that the left leg $|1\rangle$-$|4\rangle$-$|2\rangle$ is two-photon resonant. For the other leg, $|2\rangle$-$|5\rangle$-$|3\rangle$, we define $\delta_3=\delta_{30}+\Delta$ where $\Delta=0$ corresponds to the two-photon resonant condition of the right leg, while taking into account the light shift of level $|3\rangle$ caused by $\Omega_4$ [$\Omega_4^2/(4\delta_4)$], as well as the light shift of level $|2\rangle$ mentioned above. Due to the Raman-type population inversion between levels $|1\rangle$ and $|2\rangle$, $\Omega_2$ will experience Raman gain in the presence of $\Omega_1$. Similarly, $\Omega_3$ will experience Raman gain in the presence of $\Omega_4$ due to the population inversion between levels $|3\rangle$ and $|2\rangle$. However, when both legs are two photon resonant, the Raman transition amplitude from $|1\rangle$ to $|2\rangle$ can cancel that from $|3\rangle$ to $|2\rangle$, which is similar to the dark state in the electromagnetically induced transparency.

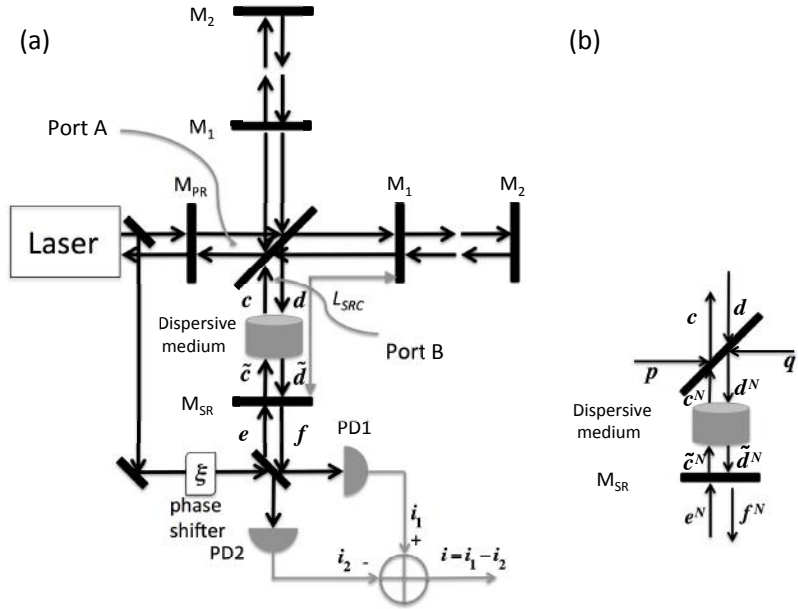

FIG. 3. (a) WLC-SR design. A dispersive medium realized using $^{87}$Rb atoms is inserted in the SR cavity. Here the input field is denoted by $e$ and the output field is denoted by $f$. (b) The SR cavity where the QN from the dispersive medium is modeled by inserting a beam splitter with power reflectivity $R_{BS}$ and transmissivity $T_{BS}$. Here $p$ and $q$ denote the vacuum noises that leak into the system.

This system can be used as the NDM in the white-light-cavity signal-recycling (WLC-SR) [8,9] interferometeric gravitational wave detector, which is shown schematically in Fig. 3. Here we use a Michelson interferometer with arm cavities and dual recycling, which is the scheme used by the advanced Laser Interferometric Gravitational wave Observatory (aLIGO) [11], and an NDM is inserted in the signal recycling (SR) cavity in order to compensate the phase variation as the frequency varies. Current LIGO operates at 1064nm, which is different from the probe frequency in our system. However, future LIGO may operate at a wavelength consistent with this system. In order to use this system in LIGO, the interferometer needs to be illuminated with a circularly ($\sigma^+$) polarized laser instead of the current linearly polarized laser, at ~795nm. The pump fields corresponding to $\Omega_1$, $\Omega_2$ and $\Omega_4$ have to be generated by splitting a part of the main laser, followed by frequency shifting using acoustic-optic modulators, for example. The optical pumping beams at ~780nm can be generated from an independent, separate laser.

### III. THEORETICAL MODEL OF THE GEIT SYSTEM

We develop a full-blown model that takes into account all the relevant Zeeman sublevels participating in the process, using the density-matrix approach. The three $\pi$-polarized fields are assumed to be resonant with the corresponding transitions and have Rabi frequencies of $\Omega_{p,\pi k}$ ($k=1,2,3$). The $\sigma^+$-polarized and $\sigma^-$-polarized fields applied on the D2 transition each couples four transitions. We express all the Rabi frequencies as multiples of the Rabi frequency for the transition with the smallest dipole moment, since the Rabi frequency of each transition is proportional to its corresponding dipole moment matrix element [12]. For example, we write the Rabi frequencies of the transitions $|6\rangle$-$|8\rangle$, $|2\rangle$-$|9\rangle$, $|7\rangle$-$|10\rangle$, and $|3\rangle$-$|11\rangle$, respectively, as:

$$\Omega_1^- = -\Omega_{p,\sigma-}, \Omega_2^- = -\sqrt{\frac{3}{2}}\Omega_{p,\sigma-}, \Omega_3^- = -\sqrt{\frac{3}{2}}\Omega_{p,\sigma-}, \Omega_4^- = -\Omega_{p,\sigma-}, \tag{0}$$

Similarly, we write the Rabi frequencies of the transitions $|7\rangle$-$|12\rangle$, $|2\rangle$-$|11\rangle$, $|6\rangle$-$|10\rangle$, and $|1\rangle$-$|9\rangle$, respectively, as:

$$\Omega_1^+ = \Omega_{p,\sigma+}, \Omega_2^+ = \sqrt{\frac{3}{2}}\Omega_{p,\sigma+}, \Omega_3^+ = \sqrt{\frac{3}{2}}\Omega_{p,\sigma-}, \Omega_4^+ = \Omega_{p,\sigma+}, \tag{0}$$

We assume that the $\sigma^-$-polarized light is resonant with the $|6\rangle$-$|8\rangle$ transition. As a result, the other three transitions are below resonance and the detunings for the four transitions are $\delta_l^- = 0.23B(l-1)\text{MHz}$, where $l = 2, 3, 4$ correspond to the $|2\rangle$-$|9\rangle$, $|7\rangle$-$|10\rangle$, and $|3\rangle$-$|11\rangle$ transitions, respectively. If we consider each transition as an effective two-level system, then the excitation at the higher level is

$$\rho_{ll}^- = \frac{(\Omega_l^-)^2}{\Gamma^2 + 2(\Omega_l^-)^2 + 4(\delta_l^-)^2}, \tag{0}$$

which is on the order of $10^{-6}$ if we take $\Omega_{p,\sigma-} \approx 0.038\text{MHz}$ and $\Gamma = 6\text{MHz}$, and decreasing with increasing value of $l$. Therefore, for simplicity, we keep the $|2\rangle$-$|9\rangle$ interactions, but neglect the $|7\rangle$-$|10\rangle$ and $|3\rangle$-$|11\rangle$ transitions (amounting to setting $\Omega_3^- = \Omega_4^- = 0$). Similarly, we assume that the $\sigma^+$-polarized light is resonant with the $|7\rangle$-$|12\rangle$ transition and neglect the couplings of the $\sigma^+$-polarized light with the $|1\rangle$-$|9\rangle$ and $|6\rangle$-$|10\rangle$ transitions (amounting to setting $\Omega_3^+ = \Omega_4^+ = 0$), while keeping the $|2\rangle$-$|11\rangle$ coupling. We will show later in this paper that the QN limited sensitivity of the WLC-SR using this system does not change significantly when the coupling of the $\sigma^-$-polarized light to the $|7\rangle$-$|10\rangle$ transition and that of the $\sigma^+$-polarized light to the $|6\rangle$-$|10\rangle$ transition are included [13].

For the fields that couple the Zeeman sublevels within the {$5S_{1/2}$, F=2} manifold to those within the {$5P_{1/2}$, F=1} manifold, there are four different laser beams with different frequencies. These beams produce additional coupling beyond those shown in Fig. 2. For example, the $\sigma^+$-polarized beam with Rabi frequency $\Omega_3$ would excite the $|1\rangle$-$|4\rangle$ transition as well. This creates a situation where the $|1\rangle$-$|4\rangle$ transition is excited simultaneously by fields at two different frequencies. Under this condition, the Hamiltonian would retain a time dependent component after the rotating wave transformation, and the resulting solution of the density matrix would have terms that are harmonics of the frequency corresponding to this time dependent term. Furthermore, similar effect would occur for the $|2\rangle$-$|4\rangle$, $|2\rangle$-$|5\rangle$, and $|3\rangle$-$|5\rangle$ transitions as well, making it exceedingly difficult to simulate the behavior of the system more exactly. To circumvent this problem, we have only considered interactions that couple the Zeeman sublevels within the {5S$_{1/2}$, F=2} manifold via two-photon resonances. Since, in steady state, most of the populations are in levels $|1\rangle$ and $|3\rangle$ within this manifold, such an approximation is justified. In the same vein, we have ignored all couplings of levels $|6\rangle$ and $|7\rangle$ caused by these four beams, since optical pumping moves atom out of these two states.

The {5P$_{1/2}$, F=2} manifold is not shown in the scheme in Fig. 2. For example, the pump field $\Omega_1$ couples to not only the transition from level $|1\rangle$ to level $|4\rangle$ in the {5P$_{1/2}$, F=1} manifold, but also the transition from level $|1\rangle$ to $|5P_{1/2}, F=2, m_F=-1\rangle$ in the {5P$_{1/2}$, F=2} manifold, with a Rabi frequency of $\Omega'_1$ and a detuning of $\delta'_1$. This additional coupling will introduce an additional light shift $\Omega'^2_1/(4\delta'_1)$ to level $|1\rangle$. It can be taken into account using an effective Rabi frequency $\tilde{\Omega}_1$ which satisfies the condition that $\tilde{\Omega}^2_1/(4\delta_1) = \Omega^2_1/(4\delta_1) + \Omega'^2_1/(4\delta'_1)$. Similarly, we can use effective Rabi frequencies for the rest of the pumps fields and the probe field $\Omega_j$ ($j=2,3,4$) to take into account the light shifts induced by the additional couplings to the {5P$_{1/2}$, F=2} manifold. Thus, the {5P$_{1/2}$, F=2} manifold can be incorporated by effective Rabi frequencies for the pump and probe fields.

The decay of the upper levels to the ground states are included using the decay rates $\Gamma_m$ ($m=4,5,8,9,10,11,12,16,17$). The time-independent Hamiltonian after the rotating wave approximation (RWA) and the rotating wave transformation can be written as (setting $\hbar=1$):

$$\tilde{H}_{1,1}=0, \ \tilde{H}_{2,2}=\delta_1-\delta_2, \ \tilde{H}_{3,3}=\delta_1-\delta_2+\delta_3-\delta_4, \ \tilde{H}_{4,4}=\delta_1-i\Gamma_4/2, \ \tilde{H}_{5,5}=\delta_1-\delta_2+\delta_3-i\Gamma_5/2, \quad (0)$$

$$\tilde{H}_{6,6}=\Delta_{B1}, \ \tilde{H}_{7,7}=5\Delta_{B1}-4\Delta_{B3}, \ \tilde{H}_{8,8}=\Delta_{B1}-i\Gamma_8/2, \ \tilde{H}_{9,9}=\delta_1-\delta_2-\Delta_{B1}+\Delta_{B3}-i\Gamma_9/2, \quad (0)$$

$$\tilde{H}_{10,10}=3\Delta_{B1}-2\Delta_{B3}-i\Gamma_{10}/2, \ \tilde{H}_{11,11}=\delta_1-\delta_2+\Delta_{B1}-\Delta_{B3}-i\Gamma_{11}/2, \ \tilde{H}_{12,12}=5\Delta_{B1}-4\Delta_{B3}-i\Gamma_{12}/2, \quad (0)$$

$$\tilde{H}_{13,13}=3\Delta_{B1}-2\Delta_{B3}-\Delta_{B4}, \ \tilde{H}_{14,14}=3\Delta_{B1}-2\Delta_{B3}, \ \tilde{H}_{15,15}=3\Delta_{B1}-2\Delta_{B3}+\Delta_{B4}, \quad (0)$$

$$\tilde{H}_{16,16}=3\Delta_{B1}-2\Delta_{B3}-\Delta_{B4}-i\Gamma_{16}/2, \ \tilde{H}_{17,17}=3\Delta_{B1}-2\Delta_{B3}+\Delta_{B4}-i\Gamma_{17}/2, \quad (0)$$

$$\tilde{H}_{1,4}=\Omega_1/2=\tilde{H}_{4,1}, \ \tilde{H}_{2,4}=\Omega_2/2=\tilde{H}_{4,2}, \ \tilde{H}_{2,5}=\Omega_3/2=\tilde{H}_{5,2}, \ \tilde{H}_{3,5}=\Omega_4/2=\tilde{H}_{5,3}, \quad (0)$$

$$\tilde{H}_{10,14}=\Omega_{p,\pi 3}/2=\tilde{H}_{14,10}, \ \tilde{H}_{13,16}=\Omega_{p,\pi 1}/2=\tilde{H}_{16,13}, \ \tilde{H}_{15,17}=\Omega_{p,\pi 2}/2=\tilde{H}_{17,15}, \quad (0)$$

$$\tilde{H}_{6,8}=\tilde{H}_{8,6}=\Omega^-_1/2, \tilde{H}_{2,9}=\tilde{H}_{9,2}=\Omega^-_2/2, \tilde{H}_{7,10}=\tilde{H}_{10,7}=\Omega^-_3/2, \tilde{H}_{3,11}=\tilde{H}_{11,3}=\Omega^-_4/2, \quad (0)$$

$$\tilde{H}_{7,12}=\tilde{H}_{12,7}=\Omega^+_1/2, \tilde{H}_{2,11}=\tilde{H}_{11,2}=\Omega^+_2/2, \tilde{H}_{6,10}=\tilde{H}_{10,6}=\Omega^+_3/2, \tilde{H}_{1,9}=\tilde{H}_{9,1}=\Omega^+_4/2, \quad (0)$$

where $\Delta_{B1} = (0.7B)\text{MHz}$, $\Delta_{B2} = (-0.23B)\text{MHz}$, $\Delta_{B3} = (0.93B)\text{MHz}$, $\Delta_{B4} = (-0.7B)\text{MHz}$, and $\Delta_{B5} = (0.93B)\text{MHz}$ are the Zeeman splitting between adjacent $m_F$ levels in the five hyperfine levels $\{5S_{1/2}, F=2\}$, $\{5P_{1/2}, F=1\}$, $\{5P_{3/2}, F=2\}$, $\{5S_{1/2}, F=1\}$, and $\{5P_{3/2}, F=1\}$, respectively. The remaining terms of $\tilde{H}$ are all zero. The equation of evolution for the density operator can be expressed as

$$\frac{\partial \tilde{\rho}}{\partial t} = -\frac{i}{\hbar}(\tilde{H}\tilde{\rho} - \tilde{\rho}\tilde{H}^*) + \left(\frac{\partial \tilde{\rho}}{\partial t}\right)_{source}, \quad (0)$$

where the second term represents the influx of atoms into a state due to decay from another state [14]. The decay rates between any two Zeeman sub-levels are proportional to the squares of the dipole moment matrix elements. As a result, the source terms are expressed as

$$\left(\frac{\partial \tilde{\rho}_{1,1}}{\partial t}\right)_{source} = \frac{\Gamma_4}{2}\tilde{\rho}_{4,4} + \frac{\Gamma_8}{3}\tilde{\rho}_{8,8} + \frac{\Gamma_9}{6}\tilde{\rho}_{9,9} + \frac{\Gamma_{16}}{10}\tilde{\rho}_{16,16}, \quad (0)$$

$$\left(\frac{\partial \tilde{\rho}_{2,2}}{\partial t}\right)_{source} = \frac{\Gamma_4}{12}\tilde{\rho}_{4,4} + \frac{\Gamma_5}{12}\tilde{\rho}_{5,5} + \frac{\Gamma_9}{4}\tilde{\rho}_{9,9} + \frac{\Gamma_{11}}{4}\tilde{\rho}_{11,11} + \frac{\Gamma_{16}}{60}\tilde{\rho}_{16,16} + \frac{\Gamma_{17}}{60}\tilde{\rho}_{17,17}, \quad (0)$$

$$\left(\frac{\partial \tilde{\rho}_{3,3}}{\partial t}\right)_{source} = \frac{\Gamma_5}{2}\tilde{\rho}_{5,5} + \frac{\Gamma_{11}}{6}\tilde{\rho}_{11,11} + \frac{\Gamma_{12}}{3}\tilde{\rho}_{12,12} + \frac{\Gamma_{17}}{10}\tilde{\rho}_{17,17}, \quad (0)$$

$$\left(\frac{\partial \tilde{\rho}_{6,6}}{\partial t}\right)_{source} = \frac{\Gamma_4}{4}\tilde{\rho}_{4,4} + \frac{\Gamma_8}{6}\tilde{\rho}_{8,8} + \frac{\Gamma_9}{12}\tilde{\rho}_{9,9} + \frac{\Gamma_{10}}{4}\tilde{\rho}_{10,10} + \frac{\Gamma_{16}}{20}\tilde{\rho}_{16,16}, \quad (0)$$

$$\left(\frac{\partial \tilde{\rho}_{7,7}}{\partial t}\right)_{source} = \frac{\Gamma_5}{4}\tilde{\rho}_{5,5} + \frac{\Gamma_{10}}{4}\tilde{\rho}_{10,10} + \frac{\Gamma_{11}}{12}\tilde{\rho}_{11,11} + \frac{\Gamma_{12}}{6}\tilde{\rho}_{12,12} + \frac{\Gamma_{17}}{20}\tilde{\rho}_{17,17}, \quad (0)$$

$$\left(\frac{\partial \tilde{\rho}_{13,13}}{\partial t}\right)_{source} = \frac{\Gamma_4}{12}\tilde{\rho}_{4,4} + \frac{\Gamma_8}{2}\tilde{\rho}_{8,8} + \frac{\Gamma_9}{4}\tilde{\rho}_{9,9} + \frac{\Gamma_{10}}{12}\tilde{\rho}_{10,10} + \frac{5}{12}\Gamma_{16}\tilde{\rho}_{16,16}, \quad (0)$$

$$\left(\frac{\partial \tilde{\rho}_{14,14}}{\partial t}\right)_{source} = \frac{\Gamma_4}{12}\tilde{\rho}_{4,4} + \frac{\Gamma_5}{12}\tilde{\rho}_{5,5} + \frac{\Gamma_9}{4}\tilde{\rho}_{9,9} + \frac{\Gamma_{10}}{3}\tilde{\rho}_{10,10} + \frac{\Gamma_{11}}{4}\tilde{\rho}_{11,11} + \frac{5}{12}\Gamma_{16}\tilde{\rho}_{16,16} + \frac{5}{12}\Gamma_{17}\tilde{\rho}_{17,17}, \quad (0)$$

$$\left(\frac{\partial \tilde{\rho}_{15,15}}{\partial t}\right)_{source} = \frac{\Gamma_5}{12}\tilde{\rho}_{5,5} + \frac{\Gamma_{10}}{12}\tilde{\rho}_{10,10} + \frac{\Gamma_{11}}{4}\tilde{\rho}_{11,11} + \frac{\Gamma_{12}}{2}\tilde{\rho}_{12,12} + \frac{5}{12}\Gamma_{17}\tilde{\rho}_{17,17}. \quad (0)$$

These equations are solved in steady state. Scanning the probe detuning $\delta_3$, we plot in Fig. 4 the real and imaginary parts of the susceptibility for the probe field $\Omega_3$ normalized by the number density $n$, which is $\chi/n = -\hbar c \Gamma_{52}^2 \tilde{\rho}_{52}/(I_{sat}\Omega_3)$. Here the saturation intensity is $I_{sat} = I_{cyc}(d_{cyc}/d_{2-5})^2$, where $I_{cyc} = 16.6933\text{W/m}^2$ is the saturation intensity for the cycling transition $|F=2, m_F=2\rangle \rightarrow |F=3, m_F=3\rangle$, and $d_{cyc}$ and $d_{2-5}$ are the matrix elements for the cycling transition and the $|2\rangle \rightarrow |5\rangle$ transition, respectively. A magnetic field of 40 Gauss is used. The Rabi frequencies of the pump fields are $\Omega_1 = 0.6\text{MHz}$, $\Omega_2 = 18\text{MHz}$, and $\Omega_4 = 12\text{MHz}$, respectively, and the Rabi frequency of the probe field is $\Omega_3 = 60\text{Hz}$. The detunings of the pump fields and the probe field are all set to be ~2.23GHz. The Rabi frequencies of the optical pumping beams are $\Omega_{p,\pi 1} = \Omega_{p,\pi 2} = \Omega_{p,\pi 3} = 0.06\text{MHz}$ and $\Omega_{p,\sigma-} = \Omega_{p,\sigma+} \approx 0.038\text{MHz}$, respectively. The decay rate of the upper levels is assumed to be $\Gamma_m = 6\text{MHz}$. We show that a negative dispersion is produced in Fig. 4(a) and the transmission profile with a dip on top of a broad gain is plotted in Fig. 4(b).

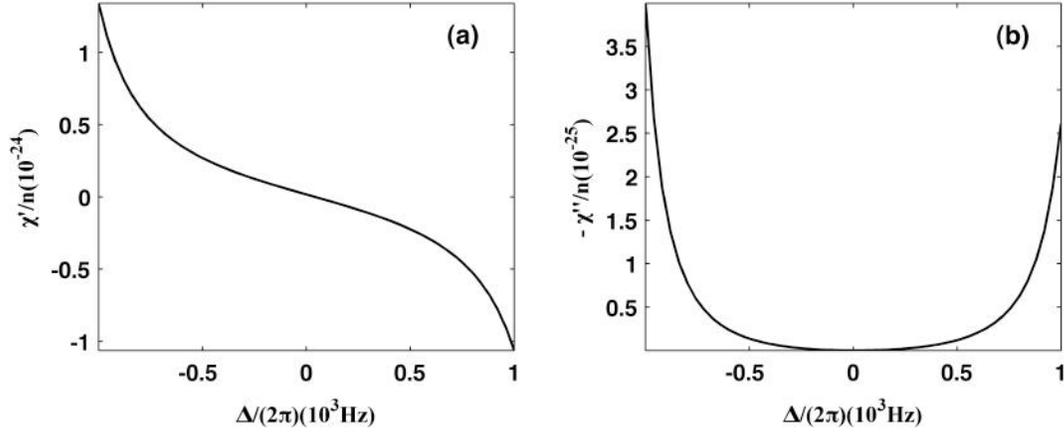

FIG. 4. (a) Imaginary and (b) real parts of $\chi / n$ ($\chi$ is the susceptibility and $n$ is the number density of Rb atoms) as a function of detuning $\Delta = \omega - \omega_0$ in the $^{87}$Rb based GEIT system. Here $\omega_0$ is the frequency corresponding to the dip in the transmission.

The QN limited sensitivity for the GW signal in the WLC-SR scheme in Fig. 3 is calculated using the input-output relation between the principal noise input $e = (e_1, e_2)^T$ from Port B and the signal and noise output $f = (f_1, f_2)^T$ [8]. Here we follow the two-photon formalism developed by Caves and Schumaker [15,16] to represent the fields as the amplitudes of the two-photon modes. In order to calculate the QN in this GEIT system accurately, the ME approach should be used [9]. However, since a total of 17 energy levels are involved, applying the full ME approach to this system is very complex. We also notice that the GEIT system is a phase-insensitive linear amplifier, and that in some phase-insensitive linear amplifiers or attenuators, the ME approach agrees closely with the Caves model [17,18], as we have also shown earlier in Ref. 8. In one of the two cases of the five-level GEIT system considered in Ref. 8 where we get an enhancement factor of 17.66 in the sensitivity-bandwidth product, the results predicted by ME and Caves model differ by ~11%, while in the other case where the enhancement factor is 16.55 the results differ by less than 0.2%. Thus, it is reasonable to assume that the enhancement in the sensitivity bandwidth product calculated by using the Caves model for a GEIT system is likely to be over-estimated by a factor on the order ranging from 1.002 to 1.11. Of course, there is no general proof for such a bound on the over-estimation. In the future, we will carry out a full-blown calculation of the QN in this system using the ME approach. For now, we proceed with the assumption stated above, and we use the Caves model to determine the QN due to the amplification from the GEIT system. We model the QN by placing inside the WLC a beam splitter [BS, as shown in Fig. 3(b)] that has a power reflectivity $T_{BS} = g$ and power transmissivity of $R_{BS} = |g-1|$, from which the vacuum fields can leak into the system from the outside. We write the input-output relation for the BS as

$$c = \sqrt{T_{BS}}c^N + \sqrt{R_{BS}}p, d^N = \sqrt{T_{BS}}d - \sqrt{R_{BS}}q,  \tag{0}$$

Using the same method as in Sec. III in Ref. 8, we plot the resulting QN curves of the WLC-SR scheme in Fig. 5, which shows an enhancement in sensitivity-bandwidth product by a factor of ~19 compared to the curve for the GW detector with signal recycling (SR) configuration with the highest sensitivity (shown as red dashed curve). If we employ the upper bound (1.11) of the overestimation factor, then the actual enhancement factor would be ~17. Here we use the density-length product of $nl = 1.25 \times 10^{18} m^{-2}$ ($n$ is the number density of Rb atoms and $l$ is the length of the NDM). For $l = 1m$, the number density required is $n = 1.25 \times 10^{18} m^{-3}$. Such a number density is achievable using a holographically shaped dark spontaneous-force optical trap (SPOT) for a dark core radius of around 6mm [19], and an atom cloud diameter of about 0.6mm. Due to the physical constraints imposed by the trap geometry, these clouds cannot be placed right next to each other. Instead, one can place the traps apart with a separation of, for example, 6cm between adjacent ones. This will represent a filling factor of 1%. Thus, in order to reach the effective value of $nl$ noted above, the actual length covered by the array of traps has to be ~100m. In order to accommodate this configuration, the distance between the beam splitter and $M_{SR}$ (in Fig. 3) has to be increased to a value of ~100m. We have verified that such a change in the relative position of the SR mirror does not affect the dynamics of the WLC-SR scheme to any noticeable degree. Finally, note that the size of the signal beam must be small than ~0.6mm at the trap location. To accommodate this constraint, the dark port output of the beam splitter will first be reduced to a diameter of ~0.4mm using a telescope. In addition, periodic refocusing lenses would be inserted, within the ~100m propagation path, to ensure that the beam diameter does not exceed ~0.5mm. While it would be a challenging task to implement such a system experimentally, it is by no means implausible.

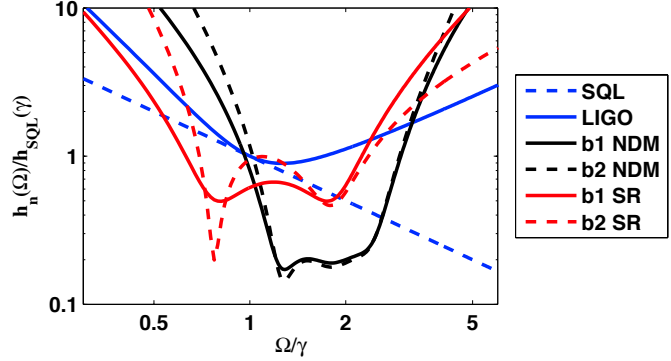

FIG. 5. Log-log plot of the normalized quantum noise $h_n(\Omega)/h_{SQL}(\gamma)$ of the GW detector versus $\Omega/\gamma$ for the first quadrature b1 and second quadrature b2, following the two-photon formalism developed in Refs. 15 and 16. Here $h_n(\Omega)$ is the square root of the noise spectral density for the GW signal at a sideband frequency $\Omega$, $h_{SQL}(\gamma)$ is the standard quantum limit for GW detection at a sideband frequency $\Omega = \gamma$, where $\gamma$ is the half bandwidth of the arm cavity of the detector. The black curves represent the quantum noise for the WLC-SR using the GEIT system as the NDM. The red curves represent the quantum noise for the GW detector with SR. The noise curve for LIGO and the standard quantum limit (SQL) curve are plotted in blue. For additional details underlying the notations used here, see, for example, Ref. 8 or Ref. 10. The noise curves for the WLC-SR scheme shows an enhancement in sensitivity-bandwidth product by a factor of ~19 compared to the curve for the SR configuration with the highest sensitivity (b2 quadrature). Since the application of the Caves model to GEIT implies an overestimation of the enhancement factor by as much as 11%, the lower bound on the enhancement factor is ~17.

Next, we consider the case where the coupling of the $\sigma^-$-polarized light to the $|7\rangle$-$|10\rangle$ transition and that of the $\sigma^+$-polarized light to the $|6\rangle$-$|10\rangle$ transition are included by taking

$$\Omega_3^- = -\sqrt{\frac{3}{2}}\Omega_{p,\sigma-}, \quad \Omega_3^+ = \sqrt{\frac{3}{2}}\Omega_{p,\sigma-}. \tag{0}$$

For this case, we again calculate the resulting dispersion and the quantum noise curves for the WLC-SR scheme. As a comparison, we show in Fig. 6 the sensitivity curves for the first quadrature, with and without taking into account the above couplings, respectively. In this case, the difference is very small. Similar agreement is seen for the second quadrature as well (not shown). This result justifies our assumption that the $|7\rangle$-$|10\rangle$ and the $|6\rangle$-$|10\rangle$ coupling can be neglected. In addition, it justifies the assumption that the $|1\rangle$-$|9\rangle$ and the $|3\rangle$-$|11\rangle$ coupling can also be neglected.

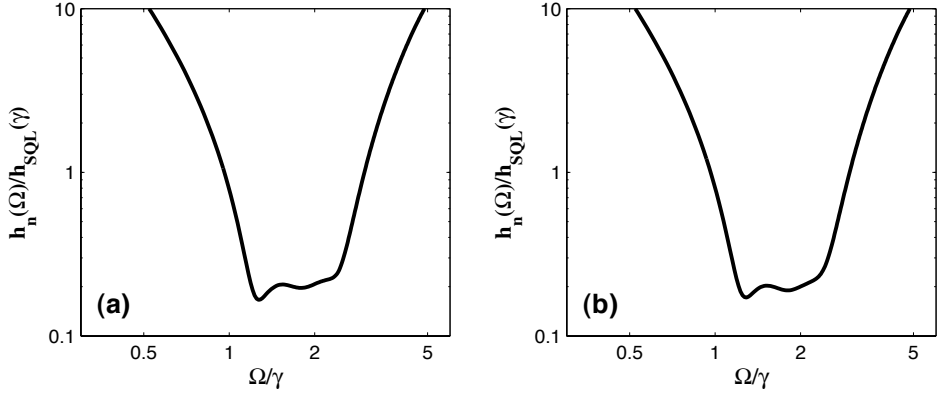

FIG. 6. Log-log plots of the normalized quantum noise $h_n(\Omega)/h_{SQL}(\gamma)$ versus $\Omega/\gamma$ for the first quadrature b1 for the WLC-SR scheme using the GEIT system with (a) and without (b) taking into account the coupling of the $\sigma^-$-polarized light to the $|7\rangle$-$|10\rangle$ transition and that of the $\sigma^+$-polarized light to the $|6\rangle$-$|10\rangle$ transition.

## IV. CONCLUSION AND FUTURE PLAN

We have presented an explicit realization of the five-level GEIT system, which shows a negative dispersion and also an EIT dip superimposed on a broad gain profile, using non-degenerate Zeeman sublevels in $^{87}$Rb atoms centered around 795nm. The current LIGO operates at 1064nm but future LIGO may operator at a wavelength that is consistent with this atomic system. Moreover, the interferometer needs to be illuminated with a circularly ($\sigma^+$) polarized laser instead of the current linearly polarized laser used in aLIGO. In that case, the GEIT system can in principle be incorporated as the negative dispersion medium in the WLC-SR scheme of gravitational wave detector. Estimating its quantum noise using the Caves model and considering an overestimation of the enhancement factor by as mush as ~11%, we determine the quantum-noise-limited enhancement in the sensitivity-bandwidth to be ~17.

In the future, we will realize this GEIT system experimentally, and verify that its gain and dispersion profile agrees with the theoretical prediction. We will also need to use the full master equation approach to calculate rigorously the quantum noise for a field interacting with this system for determining the sensitivity of the WLC-SR scheme.

## ACKNOWLEDGEMENTS

This work was supported by DARPA through the slow light program under Grant No. FA9550-07-C-0030 and by AFOSR under Grants No. FA9550-10-01-0228 and No. FA9550-09-01-682-0652.